\documentclass[twocolumn, a4paper]{revtex4}
\usepackage{graphicx}
\usepackage{dcolumn}
\usepackage{amsmath}
\begin{document}
\hsize\textwidth\columnwidth\hsize\csname@twocolumnfalse\endcsname

\title{Phenomenological Model for the 0.7 Conductance Feature in Quantum Wires}
\author{D. J. Reilly$^*$}
\affiliation{Centre for Quantum Computer Technology, School of Physics, University of New South Wales,
Sydney 2052, Australia}

\begin{abstract}
One dimensional (1D) quantum wires exhibit a conductance feature near $0.7 \times 2e^2/h$ in connection with 
many-body interactions involving the electron spin. With the possibility of exploiting this effect for novel spintronic device applications,
efforts have focused on uncovering a complete microscopic theory to explain this conductance anomaly. Here we present conductance
calculations based on a simple phenomenological model for a gate-dependent spin gap that are in excellent agreement with experimental data taken on 
ultra-low-disorder quantum wires. Taken together the phenomenology and experimental data indicate that the $0.7$ feature depends strongly on 
the potential profile of the contact region, where the reservoirs meet the 1D wire. Microscopic explanations that may 
under-pin the phenomenological description are also discussed. 
\end{abstract}

\maketitle

% Intro
The quantization of conductance in ballistic quantum wires (QWs) forms one of the cornerstones of mesoscopic physics 
\cite{vanWees_QPC,Wharam_QPC1st}. A prominent and controversial exception to this well understood phenomena is the conductance feature occurring between 
$0.5 - 0.7 \times 2e^2/h$, below the first conductance plateau, which has been observed in several different one dimensional 
(1D) systems \cite{Thomas_spin1st,KristensenPRB,Reilly_PRB,Cronenwett_07,depicciotto,morimotto,bird,Biercuk}. 
Strong evidence, initially uncovered by Thomas {\it et al.,}  \cite{Thomas_spin1st} has linked the 
occurrence of this feature (and higher order features near $1.7 \times 2e^2/h$) with many-body 
interactions involving the electron spin. Driven by the possibility of exploiting this effect for device 
applications based on the spin degree of freedom, efforts continue to focus on uncovering a detailed microscopic 
explanation for the origin of the conductance feature \cite{Meir,Cornaglia,Hirose,Matveev,Seelig,Berggren_03}.   
The work presented here shows that a simple phenomenological model for the $0.7$ conductance anomaly \cite{Reilly_PRL}  
is in excellent agreement with {\it all} of our data taken on ultra-low-disorder QWs. Motivated by the remarkable agreement 
between model and experiment, we discuss several microscopic descriptions that could account for the phenomenology.
In addition, evidence is presented linking the conductance feature to the relative 
potential mis-match between the 1D QW and the two-dimensional (2D) contact reservoirs. 

%Figure 1
\begin{figure}[t!]
\begin{center}
\includegraphics[width=8.5cm]{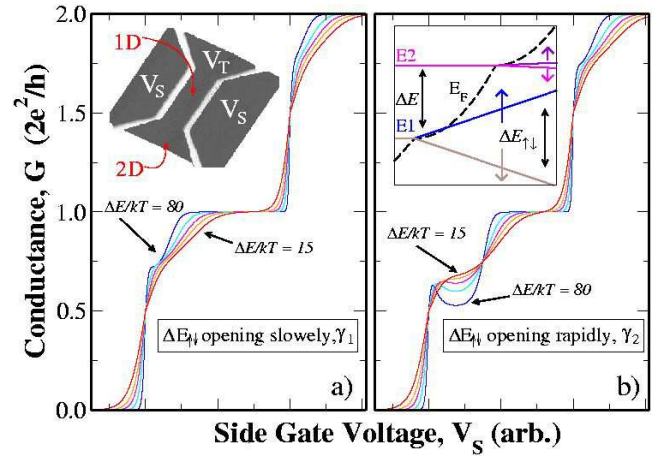}
\caption{Conductance calculations based on the model. In (a) $\gamma = d\Delta E_{\uparrow\downarrow}/dV_S$ is small in comparison to (b). 
Low temperature is shown in blue
($\Delta E/kT = 80$) and high temperature in red ($\Delta E/kT = 15$). Inset to (a) is an AFM image 
of a QW device showing the 1D and 2D regions. $V_T$ and $V_S$ are the top gate and side gates respectively.
Inset to (b) is a schematic of the model showing the Fermi level $E_F$ and the spin gap $\Delta
E_{\uparrow\downarrow}$ opening with gate bias, $V_S$.}
\vspace{-0.5cm}
\end{center}
\end{figure}

Extending our earlier work \cite{Reilly_PRL}, the phenomenological description is as follows (see inset to Fig. 1(b)). Near pinch-off, at very low gate bias the 
probability 
of transmission is equal for both spin-up and spin-down electrons. Our premise is that  
with increasing gate bias $V_S$ an energy gap forms between
up and down spins (or triplet and singlet states) and increases near linearly with 1D density $n_{1D}$. For the moment we defer discussion of the 
possible microscopic explanations for 
this gate dependent spin gap, and focus just on the phenomenology. Key to our model the Fermi-level $E_F = \hbar^2 k_F^2/2m^*$, where $m^*$ is 
the electron 
effective mass and $k_F$ is the Fermi wave vector, is parabolic with density $n_{1D}$ 
or gate bias $V_S$  since $k_F = (\pi/2)n_{1D} = (\pi/2)(c V_S/e)$, where $c$ is the capacitance between the gate and 1D electrons. 
Consistent with experimental results \cite{Reilly_PRB,Reilly_PRL,newthomas,Pyshkin}, at low temperatures this model predicts a feature near $0.5 
\times 2e^2/h$ when 
$E_F$ exceeds the spin-down energy but is yet to 
cross the spin-up band edge. As the temperature is increased the occurrence of a feature closer to $0.7 \times 2e^2/h$ is due to the continued opening of 
the spin-gap with  increasing $E_F$ so that the contribution to the current from the thermally excited electrons into the upper-spin band remains approximately
constant over a small range in $V_S$.  Although similar in spirit to the model of Bruus {\it et al.,} \cite{Bruus1} our picture is 
based on a spin-gap that is not fixed, but density-dependent and in which $0.5$ and $0.7$ features {\it do not} 
co-exist. Further, in contrast to Fermi-level `pinning' \cite{Bruus1} the model discussed here suggests that the spin-gap {\it continues} to open even as $E_F$ is above 
the 
spin-up band-edge.

%Figure 2
\begin{figure}[t!]
\begin{center}
\includegraphics[width=8.5cm]{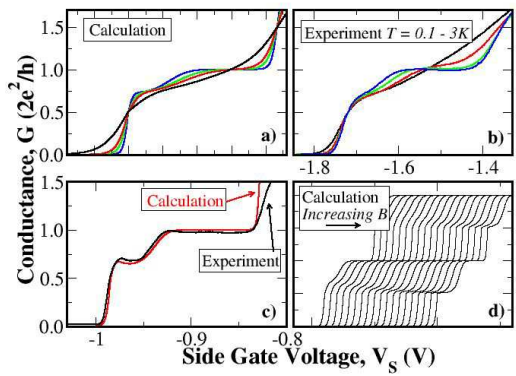}
\caption{(a) Shows the calculated temperature dependence of the $0.7$ feature in the regime of Fig. 1(a). (b) Data taken on a 
point contact device at temperatures $T$=0.5-3K (blue to black) $n_{2D}=2.1 \times 10^{11}/cm^2$. (c) Calculated 
conductance for the low temperature case of Fig. 1(b) (red) and data taken on a $l=1\mu m$ long wire at $T$=100mK, $n_{2D}=4.6 \times 10^{11}/cm^2$ (black). (d) 
shows the calculated in-plane magnetic field dependence of the $0.7$ feature. $B=0$ (left) to $B=0.1 \Delta E$ (right), traces are offset for clarity.}
\vspace{-0.5cm} 
\end{center} \end{figure}   

The only free parameter in this phenomenological model is the rate at which the spin gap $\Delta E_{\uparrow\downarrow}$ opens with gate bias $V_S$: $\gamma = d \Delta
E_{\uparrow\downarrow}/dV_S$. This rate governs the detailed shape and position of the feature as a function of temperature. Fig. 1 shows calculations based on 
this model for two different spin-gap rates, $\gamma_1 < \gamma_2$. 
The conductance is calculated in a very simple way in an effort to show the simplicity of the model. It is assumed that in the linear response regime, with a small bias 
applied between the left and right leads the conductance is approximated by:
\begin{equation}
G = 2e^2/h \int_{U_L}^{\infty} \left(-\partial f/\partial E \right) T(E) dE
\end{equation}
where $U_L$ is the bottom of the band in the left lead, $f$ is the Fermi function $f=(1/exp((E_{\uparrow\downarrow}-E_F)/kT) + 1)$ and $E_{\uparrow\downarrow}$ are
separately the spin-up and down sub-band edges.
Assuming that tunneling leads to broadening on a much smaller scale than thermal excitation, we use a classical step function for the transmission probability
$T(E) = \Theta(E_F - E_{\uparrow\downarrow})$ where $\Theta(x)=1$
for $x>E_{\uparrow\downarrow}$ and $\Theta(x)=0$ for $x<E_{\uparrow\downarrow}$.
Under this simplification  the  linear response conductance of each spin-band is well approximated by just the Fermi probability 
for thermal occupation multiplied by the conductance quantum:  
$G \sim e^2/h \times f$. 

Comparing Fig. 1(a) and 1(b) we note that the shape of the feature is 
characterized by both $\gamma = d\Delta E_{\uparrow\downarrow}/dV_S$ and $kT$ relative to the 1D sub-band spacing $\Delta E$. 
In Fig. 1(a) a feature near $0.7 \times 2e^2/h$ occurs even at low temperatures, since the spin-gap opens slowly ($\gamma_1$) as $E_F$ crosses $E_{\downarrow}$ so that 
the Fermi function also overlaps $E_{\uparrow}$ by an amount. 
Contrasting this behavior, Fig. 1(b) illustrates the regime where the spin-gap opens rapidly with $V_S$ (increased $\gamma_2$). In this case the low temperature
conductance tends towards $0.5 \times 2e^2/h$, after $E_F$ crosses $E_{\downarrow}$. Increasing the temperature causes the 
feature to broaden and rise from $0.5$ to $0.7 \times 2e^2/h$.              
 
We now turn to compare the results of our model with data taken on ultra-low-disorder QWs free from the disorder associated 
with modulation doping. Although the fabrication 
and operation of these devices has been described elsewhere \cite{BK_QWAPL}, we reiterate that they enable 
separate control of both the 2D and 1D densities (see inset Fig1. (a)). Fig. 2(a) and 2(b) compare the calculated temperature dependence of the 
$0.7$ feature to data taken on a quantum point contact device. The only parameters of the model that were adjusted are 
the sub-band energy spacing ($\Delta E$) and the rate at which the spin gap 
opens ($\gamma$) (setting an arbitrary gate capacitance $c$). As is evident, this model is in good agreement with the shape and 
dependence of the $0.7$ feature with temperature. Continuing with our comparison between model and experiment, 
Fig. 2(c) shows data taken on a QW of length $l=1\mu m$ at $T$=100mK (black) and 
calculated conductance based on the model (red), where $\gamma$ is now greater than in Fig. 2(a).
Note the non-monotonic behavior of the conductance (near $0.6 \times 2e^2/h$) which we have observed 
for many of our devices. This oscillatory structure can be traced to the parabolic dependence of $E_F$ and linear dependence of $E_{\uparrow}$ with $V_S$ in the model.    

Extending the model to include a Zeeman term: $\Delta E_{\uparrow\downarrow} \sim c V_S \pm g\mu_B B S$, where $g$ is the in-plane electron g factor, $B$ is the magnetic 
field, $\mu_B$ is the Bohr magnetron and $S = 1/2$, Fig. 2(d) shows the calculated in-plane magnetic field dependence of the $0.7$ feature. 
The calculated traces strongly resemble the experimental results of Thomas {\it et al.,} and Cronenwett {\it et al.,} \cite{Thomas_spin1st,Cronenwett_07}, in 
which the feature near $0.7 \times 2e^2/h$ evolves smoothly into the Zeeman spin-split plateau at $0.5 \times 2e^2/h$ with increasing in-plane magnetic field. 
A similar but weak dependence is also seen for the $1.7 \times 2e^2/h$ feature, where $\gamma$ has been reduced in the calculations.

The data shown in Fig. 2(c) was taken with $n_{2D}=4.6 \times 10^{11}/cm^{2}$. In comparison to Fig. 2(b) where $n_{2D}=2.1 \times 10^{11}/cm^{2}$, the high $n_{2D}$ data 
(Fig. 2(c)) shows a feature closer to $0.5 \times 2e^2/h$ and exhibits non-monotonic behavior. In the context of the model, 
$\gamma$ is the only parameter varied to achieve a fit with both the high and low  $n_{2D}$ data.  

Extending this phenomenological link between $\gamma$ and  $n_{2D}$, 
Figs. 3(a) \& 3(b) compare the model with additional data taken on a $l=1 \mu m$ QW at $T$=100mK. The different traces shown in each of the Figures correspond to 
an increasing top gate bias $V_T$ or $n_{2D}$ (right to left) for the experimental data and an 
increasing spin gap rate $\gamma = d \Delta E_{\uparrow\downarrow}/d V_g$ (right to left) for the calculations. With increasing $V_T$ (data) or $\gamma$ (calculations) 
the conductance feature exhibits an evolution from a slight shoulder feature near $0.7 \times 2e^2/h$ to a 
broader feature, approaching $0.5 \times 2e^2/h$. 

%Figure 3
\begin{figure}[t!]
\begin{center}
\includegraphics[width=8.5cm]{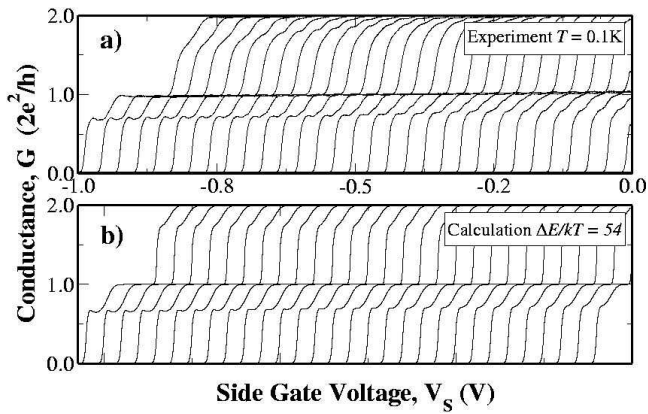}
\caption{Comparison of data taken on a $l=1 \mu m$ wire with calculations based on the model. (a) Data taken at $T$=100mK for $n_{2D}=2 - 4.6
\times 10^{11}/cm^2$ right to left. (b) Calculations for $\Delta E/kT = 54$, with $\gamma$ increasing (arb. units) right to left. Due to the 
electrostatics of the devices 
the experimental data shifts in $V_S$ with increasing $n_{2D}$ and in-turn the calculated traces have also been offset to aid in comparison with the data.}
\vspace{-0.5cm}
\end{center}
\end{figure}

The dependence of the $0.7$ feature with  $n_{2D}$ has long been debated, with different groups observing conflicting results
\cite{Thomas_int, Reilly_PRB, Nuttinck, Pyshkin, newthomas}.
We now present results that indicate that the strength and position of the feature is linked {\it not} to the absolute value of $n_{2D}$, but to the mis-match 
between the potential of the 1D and 2D regions. Fig. 4 shows data taken on a $l = 0.5 \mu m$ QW in which  $n_{2D}$ is {\it fixed}
at $n_{2D} \sim 5\times 10^{11}/cm^{2}$ and $V_{S2}$ is swept negative, reducing the conductance (see Fig. 4 inset diagram).  Traces from left to right correspond to 
$V_{S2}$ sweeps, as $V_{S1}$ is stepped more negative and $V_T$ ($n_{2D}$) is held constant.
The effect of stepping $V_{S1}$ negative is to increase the electrostatic confinement, making the relative potential difference between the 2D and 1D regions larger.
Similar to the data in Fig. 3, the feature grows in strength and lowers in conductance as $V_{S1}$ is stepped negative, although in this case $n_{2D}$ is {\it 
not} varied. Unlike the behavior expected from an impurity in the 1D channel, these results are reproducible when $V_{S1}$ and $V_{S2}$ are interchanged and the 
direction of the side-gate confinement potential is reversed. This data indicates that the position and strength of the 
$0.7$ feature depends not on the absolute value of $n_{2D}$, but the {\it relative} difference between the 1D and 2D potentials. 

%Figure 4
\begin{figure}[t!]
\begin{center}
\includegraphics[width=8.5cm]{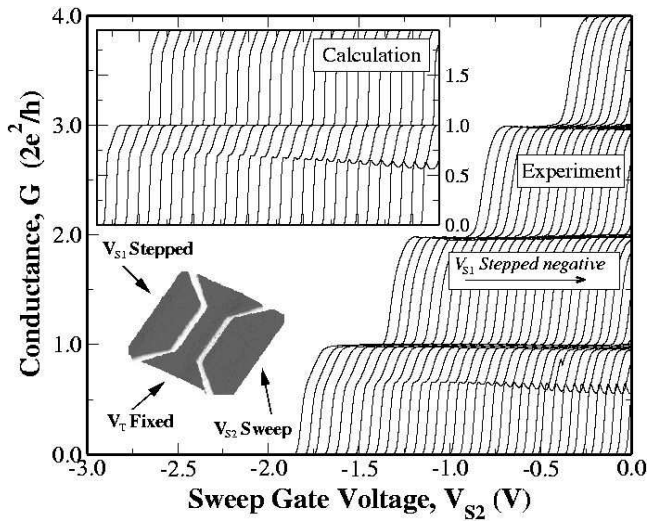}
\caption{Comparison of data taken on a $l=0.5 \mu m$ wire (right) with calculations based on the model (inset). In the experiment 
$n_{2D}$ is {\it fixed} at $\sim 5\times 10^{11}/cm^{2}$ and $V_{S2}$ is swept negative, reducing the conductance (see inset diagram for gate 
configuration). Traces from left to right correspond to $V_{S2}$ sweeps, as $V_{S1}$ is stepped more negative. For calculations 
(inset) $\gamma$ is increased in linear
steps from left to right with $\Delta E/kT = 54$ (traces off-set).}
\vspace{-0.5cm}
\end{center}
\end{figure}

Returning to the phenomenological model, we again draw a link between the 1D-2D potential profile and $\gamma$. The inset to Fig. 4 
shows calculations based on the model for differing values of $\gamma$, spanning the regime shown in the experimental results (main plot Fig. 4). 
As $\gamma$ is increased 
from left to right the feature evolves from a slight inflection to a strong non-monotonic feature consistent with the experimental data.  

Finally we compare our phenomenology with the dependence of the $0.7$ feature with applied source - drain (SD) bias. Such measurements are key 
since they permit the evolution of the 1D band-edge energies to be studied as a function of $V_S$. 
Fig. 5 compares the differential conductance ($di/dv$) of a $l=0.5\mu m$ QW (Fig. 5(a)) to calculations based on the model (Fig. 5(b)). For small 
$V_{SD}$ equation (1) for the conductance can be extended to finite $V_{SD}$, where $di/dv$ is a 
weighted average of two zero-$V_{SD}$ conductances, one for a potential 
of $E_F+\beta eV_{SD}$, and the other for $E_F-(1-\beta)eV_{SD}$, where $\beta$ characterizes 
the symmetry of the potential drop across the QW \cite{moreno}. In line with this picture Fig. 5(c) is 
a schematic showing the energies of the S and D potentials relative to the spin-band edges, in 
connection with the conductance features shown in the data and calculations. Case (1) 
corresponds to a $V_{SD}$=0 conductance of $0.6 \times 2e^2/h$ which increases to $0.8 \times 2e^2/h$ with the application
of a bias as shown in case (2). In case (3), the S and D potentials differ by one sub-band (two 
spin-bands) and the $di/dv$ exhibits the well known half-plateaus at $1.5 \times 2e^2/h$ due to the averaging of $G$ at S ($2 \times 2e^2/h$) and D ($1 \times 
2e^2/h$).  

Addressing  case (4), we focus on the $1.25 \times 2e^2/h$ features seen in the data near $V_{SD} 
\pm$8mV (Fig. 5(a)) which are mirrored in the calculations (Fig. 5(b)). To our knowledge, these features have not previously been 
discussed. In the context of our model the $1.25$ features are due to S and D 
differing by 3 spin-bands and provide evidence that the spin energy gap remains open well below the Fermi level. 
Below the first plateau a cusp feature is observed in both the data and calculations shown in Fig. 5  case (1). In the 
context of our model this cusp arises as the spin gap opens with $V_S$ so that a larger SD bias is needed before S (or D) cross $E_{\uparrow}$, and increase the 
conductance.  In regard to this cusp feature, we again note the remarkable resemblance between the experimental data and calculations based on the model. 

Having presented our model and shown it to be in excellent agreement with experimental data taken on ultra-low-disorder QWs, we now 
discuss microscopic explanations that may under-pin this phenomenology. These include spontaneous spin polarization \cite{Wang_gpar}, the Kondo effect 
\cite{Meir,Hirose,Cronenwett_07}, backscattering of electrons by acoustic phonons \cite{Seelig} and Wigner-crystallization \cite{Matveev}. The notion of a spontaneous spin 
polarization, originally suggested by Thomas {\it et al.,} \cite {Thomas_spin1st} has remained controversial in connection with exact theory 
forbidding a ferromagnetic ground state in 1D \cite{Lieb_Mattis}. 
This issue however, is complicated by the presence of 2D reservoirs that contact the 1D region and recent calculations \cite{Berggren_03} that include reservoirs suggest 
a bifurcation of ground and metastable states in association with a spin polarization. The existence of a spin-gap in connection
with such a polarized state provides a conceptual picture underlying the phenomenology presented here.      

Our phenomenology may also be consistent with a Kondo-like mechanism recently proposed to explain the $0.7$ feature \cite{Lindelof,Meir,Hirose}. In the context of a Kondo 
picture the model discussed here is suggestive of a scenario just {\it above} the Kondo temperature $T_K$, where spin screening is incomplete and a (charging) energy gap 
develops between singlet and triplet states. Recent measurements by de Picciotto {\it et al.,} \cite{depicciotto} also point to the importance of screening. 
Perhaps the dependence of $\Delta E_{\uparrow\downarrow}$ on $V_S$ and the sensitivity of the feature to the 2D-1D coupling is linked to $T_K$, 
which is a function of the hybridization energy associated with electrons tunneling from the reservoirs into the QW \cite{Meir}. Note however, that the cusp 
feature occurring at finite SD bias (discussed above), is in contrast to the Kondo-like zero-bias anomaly (ZBA) observed by Cronenwett {\it et al.,} below 
$T=$100mK \cite{Cronenwett_07}. At $T>$300mK however, the ZBA seen by Cronenwett {\it et al.,} 
evolves into a cusp feature like that seen in our data and calculations (see Fig. 2(a) in \cite{Cronenwett_07}), 
presumably due to a cross-over from $T<T_K$ to $T>T_K$. Previous investigations indicate the strength of the cusp is strongly dependent on $n_{2D}$ \cite{Reilly_PRL}. In 
this sense the absence of a ZBA in our results maybe linked to the difference in $n_{2D}$ (relative to the 1D potential) between our samples and those 
examined by Cronenwett {\it et al.,} ($n_{2D} = 1.1 \times 10^{11}/cm^2$ for Cronenwett {\it et al.,} and $n_{2D}  = 4.0 \times 10^{11}/cm^2$ for our $l=0.5\mu m$ 
wire shown in Fig. 5(a)). Such an interpretation is again consistent with $T_K$ being a function of $n_{2D}$ or the 2D-1D coupling. 
In the context of our phenomenology this implies $T_K$ is related to $\gamma$.      
%Figure 5
\begin{figure}
\begin{center}
\includegraphics[width=9.0cm]{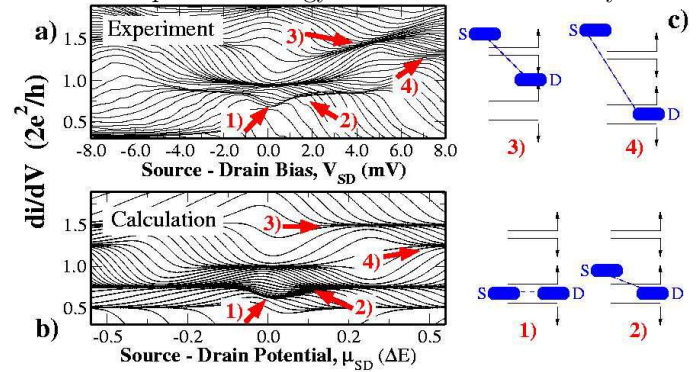}
\caption{(a) $di/dv$ data taken on a $l=0.5 \mu m$ QW at $T$=100mK. Each
trace is for a different side gate bias, $V_S$. (b) Calculations based on the phenomenological model. $di/dv$ is
plotted as a function of the difference in S-D potential in units of the sub-band spacing $\Delta E$.
(c) is a schematic showing how the positions of S and D relate to the observed
conductance features.} \vspace{-0.5cm} \end{center}
\end{figure}

Interestingly, a similar temperature dependent cross-over has been described in the theoretical work of Schmeltzer, where a short 
QW is coupled to Luttinger liquid leads \cite{Schmeltzerll} (see also \cite{Bartosch}). Further, recent work by Seelig and Matveev 
\cite{Seelig, Matveev} also describes a temperature dependent correction to the conductance and the presence of a ZBA as arising from the backscattering 
of electrons by acoustic phonons and in connection with Wigner-Crystallization. Although suggestive, further work is needed to see how these pictures 
might relate to the phenomenology discussed here. Finally we also mention that calculations based on our model (not shown) are in excellent agreement with the recent 
high-$B$ data of Graham {\it et al.,} \cite{Graham} and the shot noise measurements of Roche {\it et al.,} \cite{Roche_fano}. This agreement provides a further indication 
that our phenomenology is of general relevance and not unique to our samples or experiments.      
  
In conclusion, a phenomenological model has been shown to be in excellent agreement with data taken on ultra-low-disorder QWs. In comparing model and 
experiment, the only free parameter of the model, $\gamma$, appears to be linked to the potential mis-match between the 2D reservoirs and 1D region. 
This model provides a means of linking detailed microscopic explanations to the functional form of the $0.7 \times 2e^2/h$ 
conductance feature uncovered in experiments. Such a link is of crucial importance if this effect is to be exploited in novel spintronic devices.

This work was supported by the Australian Research Council. The Author would like to acknowledge a Hewlett-Packard Fellowship and thank
Y. Meir, K-F. Berggren, C. M. Marcus and B. I. Halperin for numerous helpful conversations and T. M. Buehler, J. L. O'Brien, A. J. Ferguson, N. J. Curson, S. Das Sarma, 
A. R. Hamilton, A. S. Dzurak and R. G. Clark for fruitful discussions.  The Author is indebted and thankful to L. N. Pfeiffer and K. W. West of Bell Laboratories for providing the excellent 
heterostructures that lead to this work.

\end{document}